\documentclass[aps,prb,twocolumn,superscriptaddress]{revtex4-2} 
\usepackage{graphicx}
\usepackage{float}
\usepackage{xcolor}
\usepackage{subfigure}
\usepackage{amsmath}
\usepackage{epsfig}
\usepackage{chemfig}
\usepackage{braket}
\usepackage{chemfig}
\usepackage{mathtools}
\usepackage{bm} 
\usepackage[colorlinks= true,urlcolor=blue,linkcolor= blue,citecolor=blue,bookmarks=false,pdfstartview=]{hyperref}
\newcommand{\CuPG}{[Cu(pyz)$_{0.5}$(gly)]ClO$_4$}

\begin{document}
\title{Asymmetric phase diagram and dimensional crossover in a system of spin-1/2 dimers under applied hydrostatic pressure}

\author{M. J. Coak}
\affiliation{Department of Physics, University of Warwick, Gibbet Hill Road, Coventry, CV4 7AL, UK}
\affiliation{London Centre for Nanotechnology, University College London, Gordon St, London, WC1H 0AH, UK}
\affiliation{School of Physics and Astronomy, University of Birmingham, Edgbaston, Birmingham, B15 1TT, UK}
\author{S. P. M. Curley}
\affiliation{Department of Physics, University of Warwick, Gibbet Hill Road, Coventry, CV4 7AL, UK}
\author{Z.~Hawkhead}
\affiliation{Department of Physics, Centre for Materials Physics, Durham University, Durham, DH1 3LE, United Kingdom}
\author{J. P. Tidey}
\affiliation{Department of Chemistry, University of Warwick, Gibbet Hill, Coventry CV4 7AL, U.K}
\author{D. Graf}
\affiliation{National High Magnetic Field Laboratory, Florida State University, Tallahassee, Florida 32310, USA}
\author{S.~J.~Clark}
\affiliation{Department of Physics, Centre for Materials Physics, Durham University, Durham, DH1 3LE, United Kingdom}
\author{P.~Sengupta}
\affiliation{School of Physical and Mathematical Sciences, Nanyang Technological University, 21 Nanyang Link, 637371 Singapore}
\author{Z. E. Manson}
\affiliation{Department of Chemistry and Biochemistry, Eastern Washington University, Cheney, Washington 99004, USA}
\author{T.~Lancaster}
\affiliation{Department of Physics, Centre for Materials Physics, Durham University, Durham, DH1 3LE, United Kingdom}
\author{P. A. Goddard}
\email{p.goddard@warwick.ac.uk}
\affiliation{Department of Physics, University of Warwick, Gibbet Hill Road, Coventry, CV4 7AL, UK}
\author{J. L. Manson}
\email{deceased 7 June 2023}
\affiliation{Department of Chemistry and Biochemistry, Eastern Washington University, Cheney, Washington 99004, USA}

\begin{abstract}
We present the magnetic and structural properties of [Cu(pyrazine)$_{0.5}$(glycine)]ClO$_4$ under applied pressure. As previously reported, at ambient pressure this material consists of quasi-two-dimensional layers of weakly coupled antiferromagnetic dimers which undergo Bose-Einstein condensation of triplet excitations between two magnetic field-induced quantum critical points (QCPs).
The molecular building blocks from which the compound is constructed give rise to exchange strengths that are considerably lower than those found in other $S = 1/2$ dimer materials, which allows us to determine the pressure evolution of the entire field-temperature magnetic phase diagram using radio-frequency magnetometry. We find that a distinct phase emerges above the upper field-induced transition at elevated pressures and also show that an additional QCP is induced at zero-field at a critical pressure of $p_{\rm c} = 15.7(5)$\,kbar. Pressure-dependent single-crystal X-ray diffraction and density functional theory calculations indicate that this QCP arises primarily from a dimensional crossover driven by an increase in the interdimer interactions between the planes. While the effect of quantum fluctuations on the lower field-induced transition is enhanced with applied pressure, quantum Monte Carlo calculations suggest that this alone cannot explain an unconventional asymmetry that develops in the phase diagram.
\end{abstract}

\maketitle

\section{Introduction}

Low-dimensional, low-spin magnetic materials in an applied magnetic field provide an unparalleled solid-state experimental testing ground for fundamental quantum phenomena, including the effects of zero-point fluctuations and quantum phase transitions~\cite{Uchiyama1999,Radu2005,Zapf2006,Coldea2010,Sachdev2011a}. In particular, the weakly coupled $S = 1/2$ antiferromagnetic (AFM) dimer system has been extensively studied and is known to exhibit a quantum disordered state, as well as Bose-Einstein condensation of triplet excitations that occurs between two field-induced quantum critical points (QCPs)~\cite{Zapf2014}. In principle, an additional QCP distinct from that manifested by applied field can be reached by enhancing the intradimer magnetic interactions using hydrostatic pressure~\cite{Rueegg2008}. This has been achieved in a few cases, however bulk magnetometry measurements are difficult to perform under high pressure and the energy scales in the materials studied so far have limited the scope of the applied-pressure measurements to only a portion of the magnetic phase diagram. Here we report magnetometry measurements at pressures up to 22\,kbar on a quasi-two-dimensional  $S = 1/2$ dimer material across a field and temperature range that spans the full magnetic phase diagram. The magnetometry results are discussed in light of pressure-dependent X-ray diffraction and density functional theory (DFT), as well as quantum Monte Carlo (QMC) calculations.

\begin{figure}[t!]
\centering
\includegraphics[width= 0.8\linewidth]{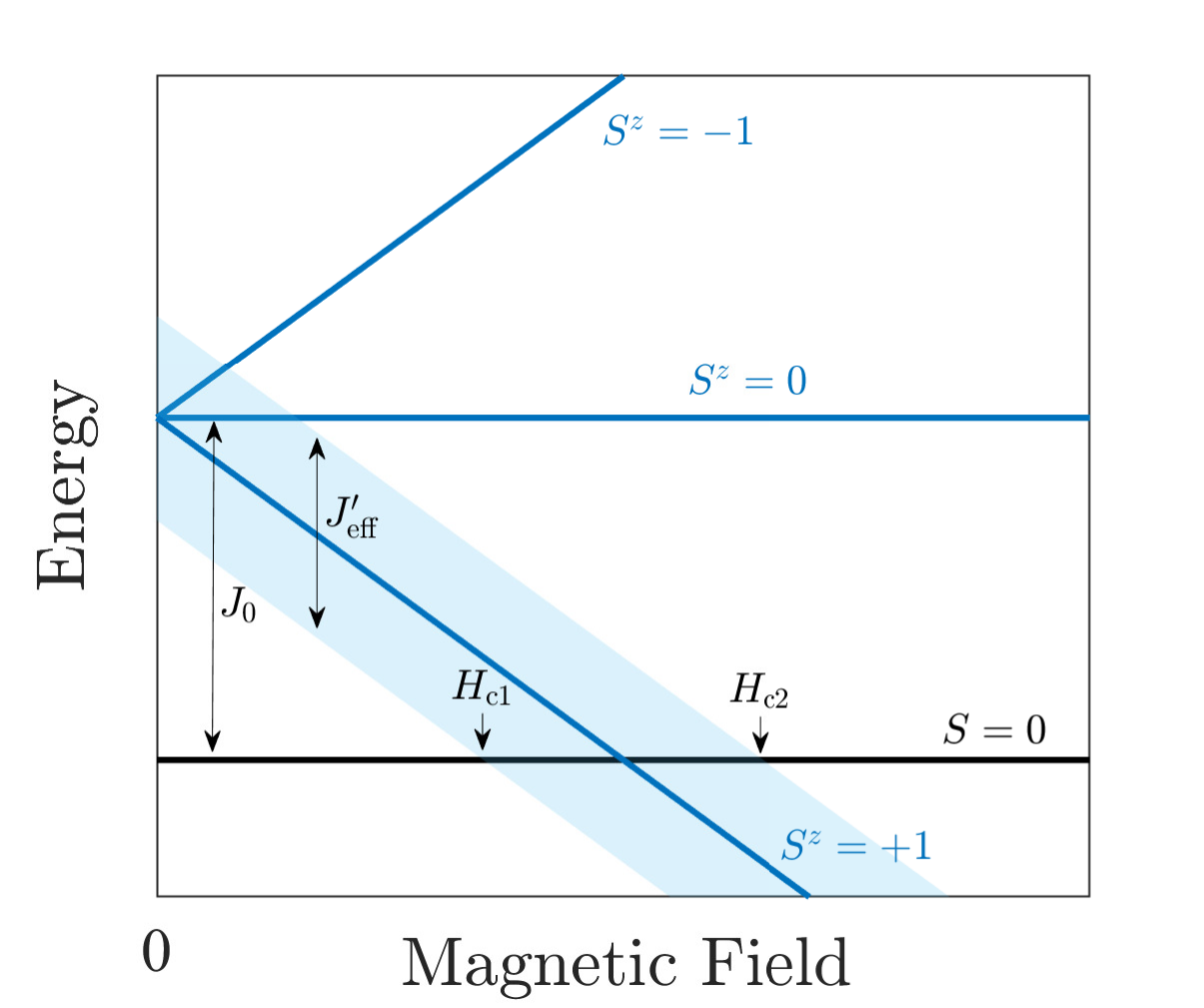}
\caption{Schematic energy level diagram for weakly coupled $S=1/2$ AFM dimers in a magnetic field.}  
\label{Fig:EnergyLevelCartoon}
\vspace{-0cm}
\end{figure}
\begin{figure*}[t!]
\centering
\includegraphics[width=\textwidth]{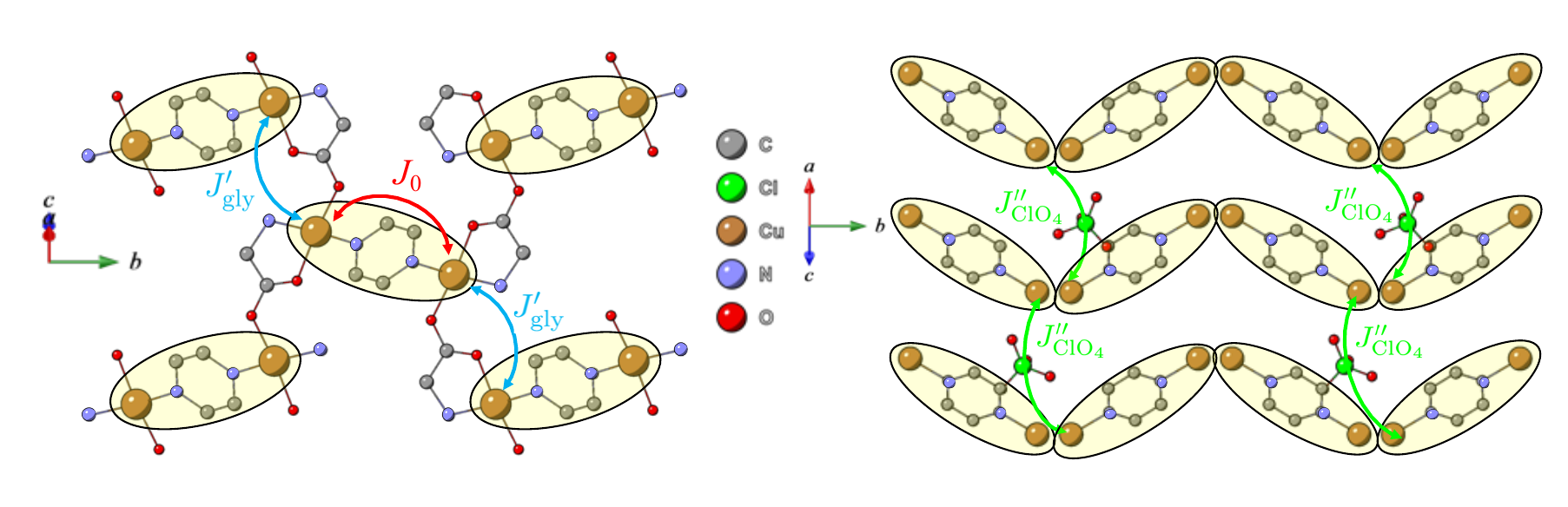}
\caption[width= \linewidth]{Ambient pressure crystal structure of [Cu(pyz)$_{0.5}$(gly)]ClO$_4$ measured at 300~K with dimer-units denoted by shaded regions. Each unit cell contains four formula units, which corresponds to two complete dimer units. (a) View along the $\left[1\,0\,-1\right]$ direction showing corrugated dimer sheets with intradimer ($J_0$) and interdimer ($J'_{\rm gly}$) exchange pathways marked with arrows. (b) View along the $\left[1\,0\,1\right]$ direction showing three layers of the corrugated dimer sheets and the  interlayer exchange ($J''_{\rm ClO_4}$) pathway between Cu(II) ions via ClO$_4$ molecules.  Adjacent dimers in each layer are connected by the glycine molecules that extend into the page, as shown in panel (a).}  \label{Fig:CuCu_struc}
\vspace{-0cm}
\end{figure*}

At low temperatures and zero magnetic field, weakly interacting Heisenberg $S = 1/2$ AFM dimers are found in a state of quantum disorder, in which zero-point fluctuations abound~\cite{Zapf2014}. The dimers are arranged as $S = 0$ singlets separated in energy from an excited degenerate $S = 1$ triplet state by an energy gap determined by the intradimer exchange energy $J_0$ (see Fig.~\ref{Fig:EnergyLevelCartoon}). Exchange pathways between dimers will give rise to an effective interdimer interaction $J'_{\rm eff}$ that acts to disperse the excited triplet, leading to a band of excitations. Turning on a magnetic field, the triplet states will split until the exchange broadened $S_z = +1$ state crosses the singlet state at the $H_{\rm c1}$ QCP and long-range $XY$-AFM order is established. Increasing the field further eventually realizes a fully field-aligned state at the $H_{\rm c2}$ QCP. The situation can be thought of as a non-magnetic vacuum from which emerge spin-1 bosonic excitations (triplets), which increase in number as field is swept, spontaneously break SO(2) symmetry to form a condensate at $H_{\rm c1}$ (provided the spin Hamiltonian itself maintains this symmetry) and reach saturation at $H_{\rm c2}$.  

The singlet-triplet gap can also be closed by tuning $J_0$ and/or $J'_{\rm eff}$ via application of pressure in the absence of a symmetry-breaking magnetic field. This QCP occurs at a pressure $p_{\rm c}$, beyond which long-range order develops that now spontaneously breaks SO(3) symmetry~\cite{Zapf2014}. While several dimer systems have been shown to exhibit a reduction in the gap with pressure, in only a few compounds has it been possible to suppress the QCP completely to zero field; namely, TlCuCl$_3$~\cite{Rueegg2004}, KCuCl$_3$~\cite{Goto2006} and (C$_4$H$_{12}$N$_2$)Cu$_2$Cl$_6$ (PHCC)~\cite{Thede2014,Bettler2017} with $p_{\rm c} \approx 1$, 8 and 4\,kbar, respectively. All of these materials possess large $H_{\rm c2}$ values beyond the range of superconducting magnets ($\approx 100$\,T for TlCuCl$_3$; 55\,T for KCuCl$_3$; and 37\,T for PHCC at ambient pressure, see~\cite{Zapf2014} and references therein), which has limited previous pressure studies to the low-field portion of the phase diagram.

Here we study the quasi-two-dimensional (Q2D) $S=1/2$ dimer system [Cu(pyz)$_{0.5}$(gly)]ClO$_4$ (pyz = pyrazine = C$_4$H$_4$N$_2$; gly = glycine = C$_2$H$_5$NO$_2$)~\cite{Lancaster2014,Brambleby2017} which, thanks to the reduced magnetic energy scales manifested in molecule-based quantum magnets, has a uniquely accessible ambient pressure $H_{\rm c2}$ close to 6\,T. Using a radio-frequency technique we are able to perform direct magnetometry measurements that encompass both field-induced QCPs up to and beyond the critical pressure of 15.7(5)\,kbar. In this way, we have found that the effect of quantum fluctuations on the low-field QCP is strongly enhanced as it is suppressed to lower fields while the high-field QCP remains largely unaffected.
We also find that a new, possibly canted, magnetic phase develops above $H_{\rm c2}$ at high pressures.

Furthermore, by performing detailed structural and DFT studies under pressure and comparing our results to QMC simulations, we deduce that the pressure-induced quantum phase transition is primarily driven by an interlayer interdimer interaction that grows with pressure while the intradimer coupling remains roughly constant, eventually causing a crossover from two to three-dimensional magnetism. An asymmetry develops in the $H$-$T$ phase diagram as pressure increases likely arising from a subtle field dependence of the secondary magnetic interactions.   

Below, after introducing the material in question and describing the experimental methods, we present the magnetometry, X-ray diffraction and DFT data. All the results are compared and their implications considered in the Discussion section.

\subsection{[Cu(pyz)$_{0.5}$(gly)]ClO$_4$}

[Cu(pyz)$_{0.5}$(gly)]ClO$_4$ crystallises into a monoclinic structure with space group $P 2_1/n$ (see Fig.~\ref{Fig:CuCu_struc})~\cite{Lancaster2014,Brambleby2017}. The material is based on a lattice of weakly interacting $S = 1/2$ dimers, with the dimer-unit itself composed of two Cu ions linked by a bridging pyz molecule which mediates the primary intradimer exchange interaction ($J_0$). The Cu-pyz-Cu dimer-unit is indicated in Fig.~\ref{Fig:CuCu_struc} with shaded regions. 

As described in detail in Refs~\cite{Lancaster2014} and \cite{Brambleby2017}, glycine ligands mediate the primary interdimer exchange ($J'_{\rm gly}$) and bridge each dimer to four nearest dimer neighbours. The dimers are arranged in corrugated sheets which stack along the [10$\overline{1}$] direction. The sheets are separated by ClO$_4$ counter ions, which bond to the dimer along the Jahn-Teller axis of the Cu. Due to this, the $d_{x^2-y^2}$ orbitals on the Cu sites lie within the equatorial CuO$_2$N$_2$ plane, such that at ambient pressure the Cu$\cdots$O---Cl---O$\cdots$Cu coordination bond mediates only very weak secondary interdimer exchange ($J''_{\rm ClO_4}$) between the layers, as shown in Fig.~\ref{Fig:CuCu_struc}(b).

As a result of this structure,  [Cu(pyz)$_{0.5}$(gly)]ClO$_4$ can be modelled as a lattice of weakly coupled $S$ = 1/2 AFM dimers interacting via Heisenberg exchange, with the magnetic properties summarised by:  
\noindent
\begin{multline}\label{eq:Hamiltonian}
\mathcal{H} = J_0\sum\limits_{i} {\bf{\hat{S}}}_{1,i}\cdot{\bf{\hat{S}}}_{2,i}
+ J'_{\rm gly}\sum\limits_{<mnij>\parallel}
{\bf{\hat{S}}}_{m,i}\cdot{\bf{\hat{S}}}_{n,j} \\
+ J''_{\rm ClO_4}\sum\limits_{<mnij>\perp}
{\bf{\hat{S}}}_{m,i}\cdot{\bf{\hat{S}}}_{n,j} - g\mu_{\rm{B}}\mu_{0}H\sum \limits_{mi} {\hat{S}}^{z}_{\textit{m,i}}
\end{multline}
\noindent
where $i$ and $j$ denote dimers and $m,n$ = 1,2 label magnetic sites within a dimer. The first term represents the intradimer interaction, while the second and third terms are the interdimer interactions between neighboring dimers within and perpendicular to the corrugated sheets, respectively. The structure also allows for an additional Dzyaloshinskii-Moriya (DM) interaction term in the Hamiltonian of the form ${\boldsymbol{D}} \cdot ({\bf{S}}_1\times {\bf{S}}_2)$, but no effects of such a perturbation have so far been detected in the temperature ranges measured and the effective spin model relevant to the experiments retains SO(2) symmetry, at least at ambient pressure~\cite{Brambleby2017}.

Previous data showed the expected lack of long-range ordering down to $T \approx 32$\,mK at zero field, the existence of two field-induced QCPs (determined from magnetization, muon spin rotation and heat capacity measurements), and a $H$-$T$ phase diagram consistent with other $S = 1/2$ dimer materials~\cite{Lancaster2014,Lancaster2018}. Reference~\cite{Brambleby2017} found that $J_0 = 5.8(3)$\,K and $J'_{\rm gly} = 1.6(1)$\,K, while $J''_{\rm ClO_4}$ is small but non-zero, with the system being well described as Q2D network of dimers with AFM coupling between the dimers. This is further supported by our DFT calculations discussed below.

\section{Methods}

\subsection{Synthesis}
Single crystal samples of [Cu(pyz)$_{0.5}$(gly)]ClO$_4$ were prepared according to the procedure established in Ref.~\cite{Lancaster2014} (and the supporting information therein).

\subsection{Crystal structure determination}
The blue crystal of dimensions 0.26 x 0.24 x 0.10 mm was loaded into an Almax-EasyLab Diacell TozerDAC alongside a ruby chip standard for pressure measurement \citep{Mao1986}. The cell employed a pre-indented steel gasket of ~150~$\mu$m thickness with a 500~$\mu$m diameter hole mechanically drilled to form the sample chamber. 4:1 MeOH/EtOH (non-dried) solution was used as the pressure transmitting medium. Data were collected at 0.5, 3.9, 7.4, 9.8, 15.5, 17.9, 21.4, 23.1, 26.6, and 28.2 kbar on a Rigaku Oxford Diffraction SuperNova diffractometer employing mirror monochromated Mo K$\alpha$ radiation from a microfocus source detected at an AtlasS2 CCD area detector. Uncertainty on these pressure values was estimated at $\pm$1~kbar from the ruby spectroscopy linewidths and differences in their positions before and after X-ray diffraction measurements.

All data were collected, indexed and integrated using CrysAlisPRO (version 1.171.43.67a (Rigaku Oxford Diffraction, 2023)), which also handled scaling and absorption. For the ambient data collection, a numerical absorption correction was performed using a face-based model of the crystal \citep{Clark1995}. For high-pressure data, absorption was handled by an empirical correction using spherical harmonics, implemented in the SCALE3 ABSPACK scaling algorithm. The ambient structure was solved by ShelXT \citep{Sheldrick2015} while high-pressure data were solved using isomorphous replacement and all models further refined using ShelXL \citep{Sheldrick2015}, with both stages performed in Olex2 version 1.5-ac6-011 \citep{Dolomanov2009}.  Complete .CIF files, including embedded structure factors and SHELX refinement instructions (.res), are made available via. the CCDC database, Deposition Numbers CCD 2302931-2302941.

\vspace{-0.5cm}
\subsection{Magnetometry}
A radio-frequency (RF) susceptometry technique was used in which the sample is placed within a small detector coil of around 30 turns that forms the inductive part of an $LCR$ circuit driven by a tunnel diode oscillator (TDO). Changes in the oscillation frequency ($\Delta\omega$) of the circuit are measured and related to the real part of the dynamic susceptibility,
\begin{equation}
    \frac{\Delta\omega}{\omega} = -\pi f \frac{{\rm d}M}{{\rm d}H}
\end{equation}
where $f$ is the filling factor of the detector coil~\cite{Clover1970,Coffey2000,Ghannadzadeh2011}.

\begin{figure*}
\centering
\includegraphics[width= 0.32\linewidth]{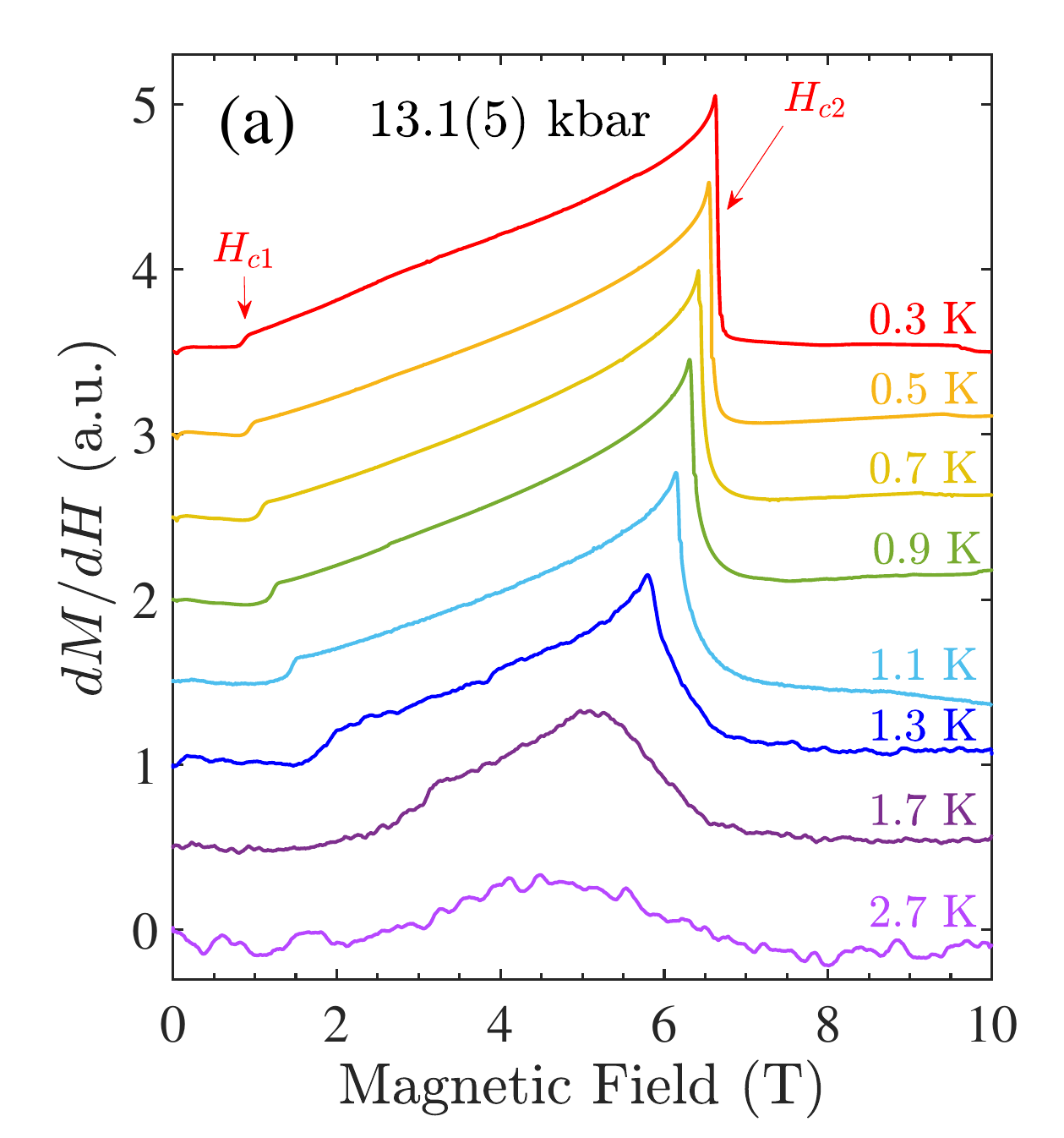}
\includegraphics[width= 0.32\linewidth]{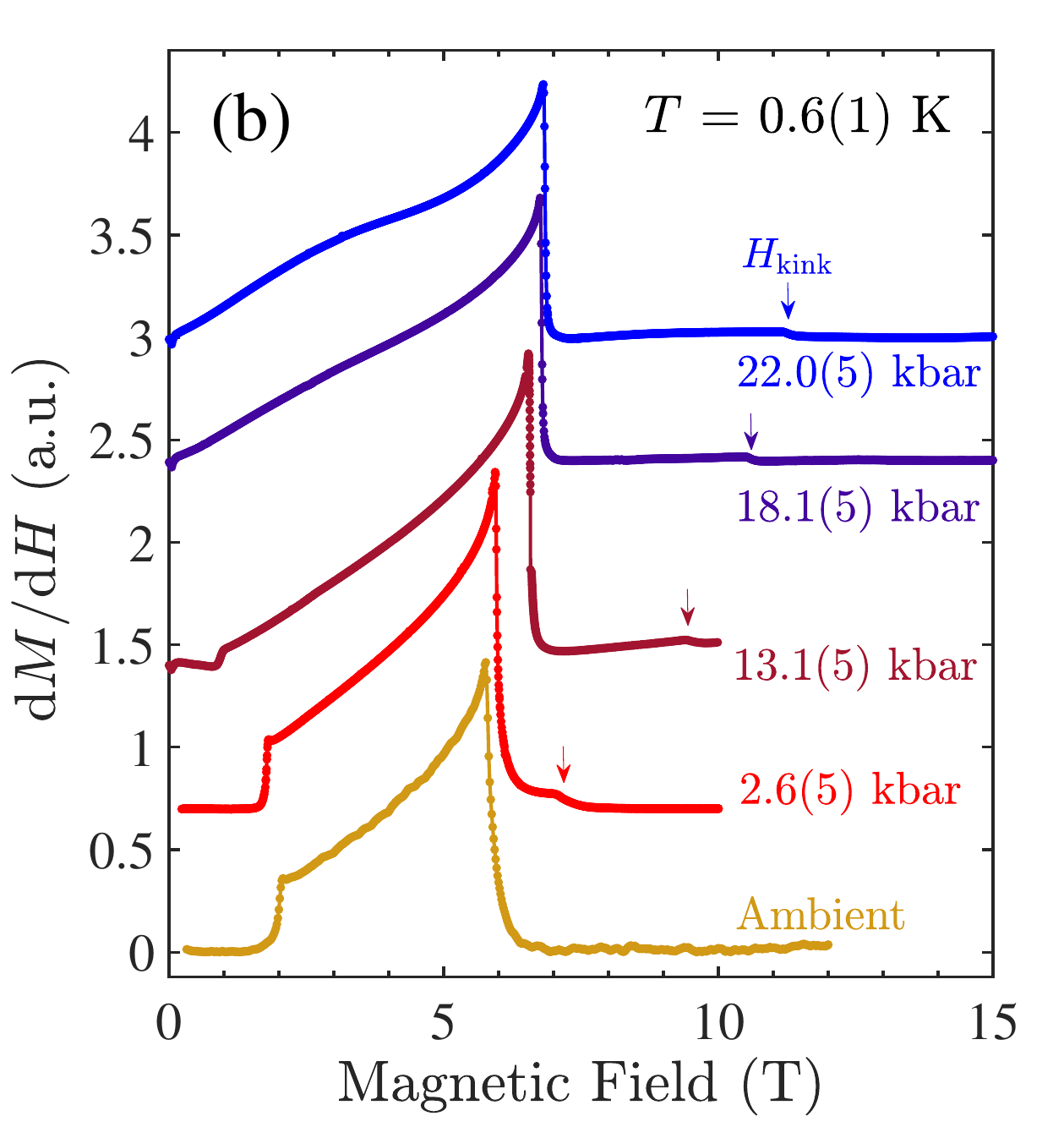}
\includegraphics[width= 0.32\linewidth]{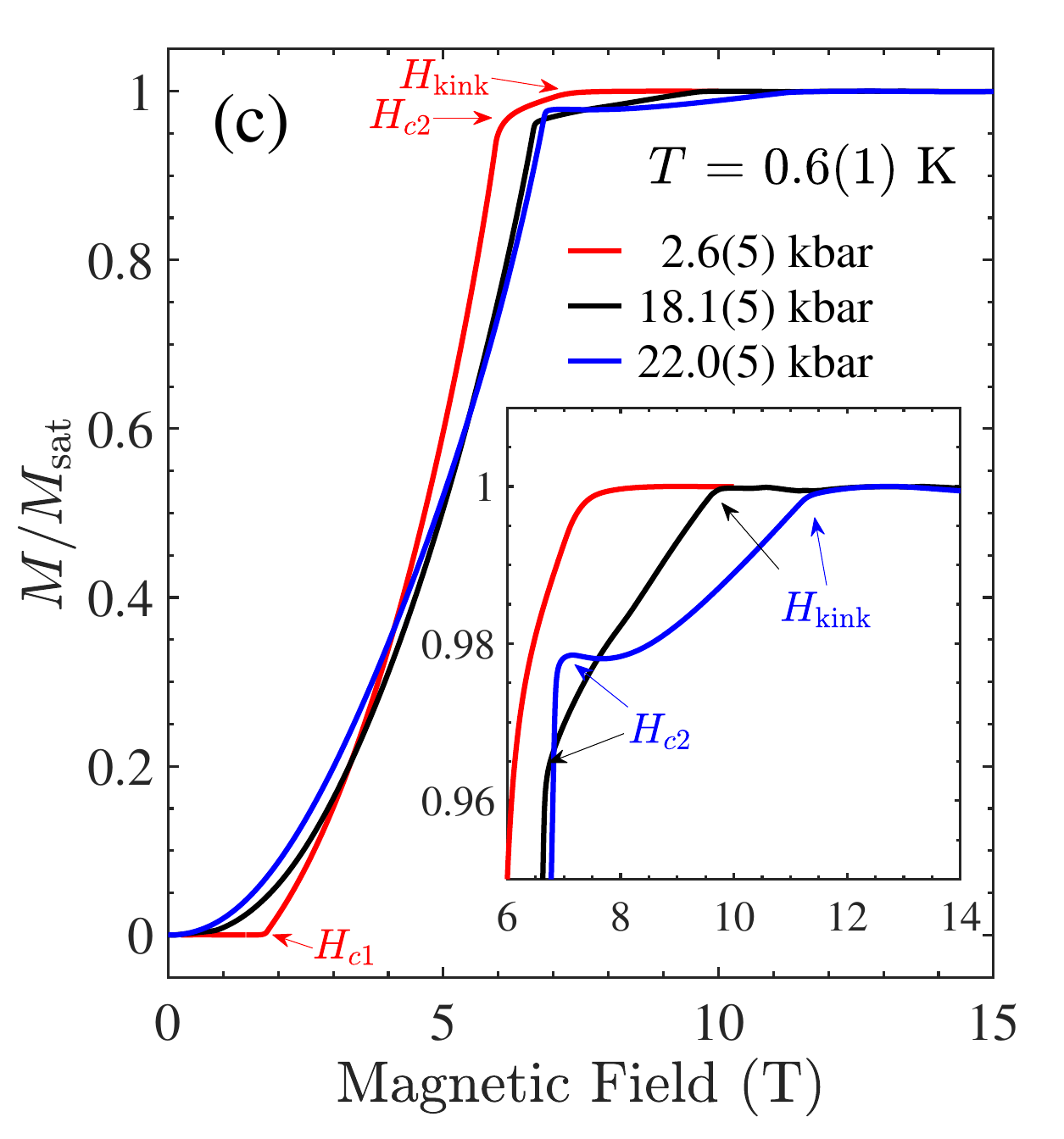}
\caption{(a) Field dependence of the dynamic susceptibility (d$M$/d$H$) measured at 13.1(5)\,kbar and various fixed temperatures for $H\parallel [011]$. (b) Pressure evolution of d$M$/d$H$ at 0.6(1)\,K. (Ambient-pressure data taken from \cite{Brambleby2017}.) (c) Pressure evolution of the magnetisation $M$($H$) measured at 0.6(1)~K determined by integrating the data in (b). The inset highlights the $H_{\rm kink}$ feature. The first critical field ($H_{\rm c1}$), second critical field ($H_{\rm c2}$) and kink-like feature ($H_{\rm kink}$) are marked with arrows.}  \label{Fig:CuCu_mag_pdep}
\vspace{-0cm}
\end{figure*}

Pressure dependence susceptometry was measured by placing the RF detector coil, containing a single crystal of [Cu(pyz)$_{0.5}$(gly)]ClO$_4$, in a piston-cylinder cell inserted in an 18\,T superconducting magnet at the National High Magnetic Field Laboratory, Tallahasee, Fl, USA. Additional measurements were taken at the University of Warwick using a pumped-$^3$He probe within the variable temperature insert of a 17\,T superconducting magnet. The field was applied parallel to the $[011]$ crystallographic direction. The pressure media used were glycerol (for X-ray diffraction measurements) and Daphne 7373 oil (for magnetometry), which give good hydrostatic pressure conditions under the pressures employed \citep{Klotz2009}. Values of pressure were determined in situ by tracking the pressure-dependence of the fluorescence spectrum of a ruby chip located on the end of a fiber-optic cable situated next to the sample. Measurements were made at 1.5, 2.6, 7.9, 13.1, 15.4, 18.1, 20.8 and 22\,kbar with an uncertainty of 0.5\,kbar estimated from the ruby fluorescence line widths.

\begin{figure*}[t]
\centering
\includegraphics[width= 0.98\linewidth]{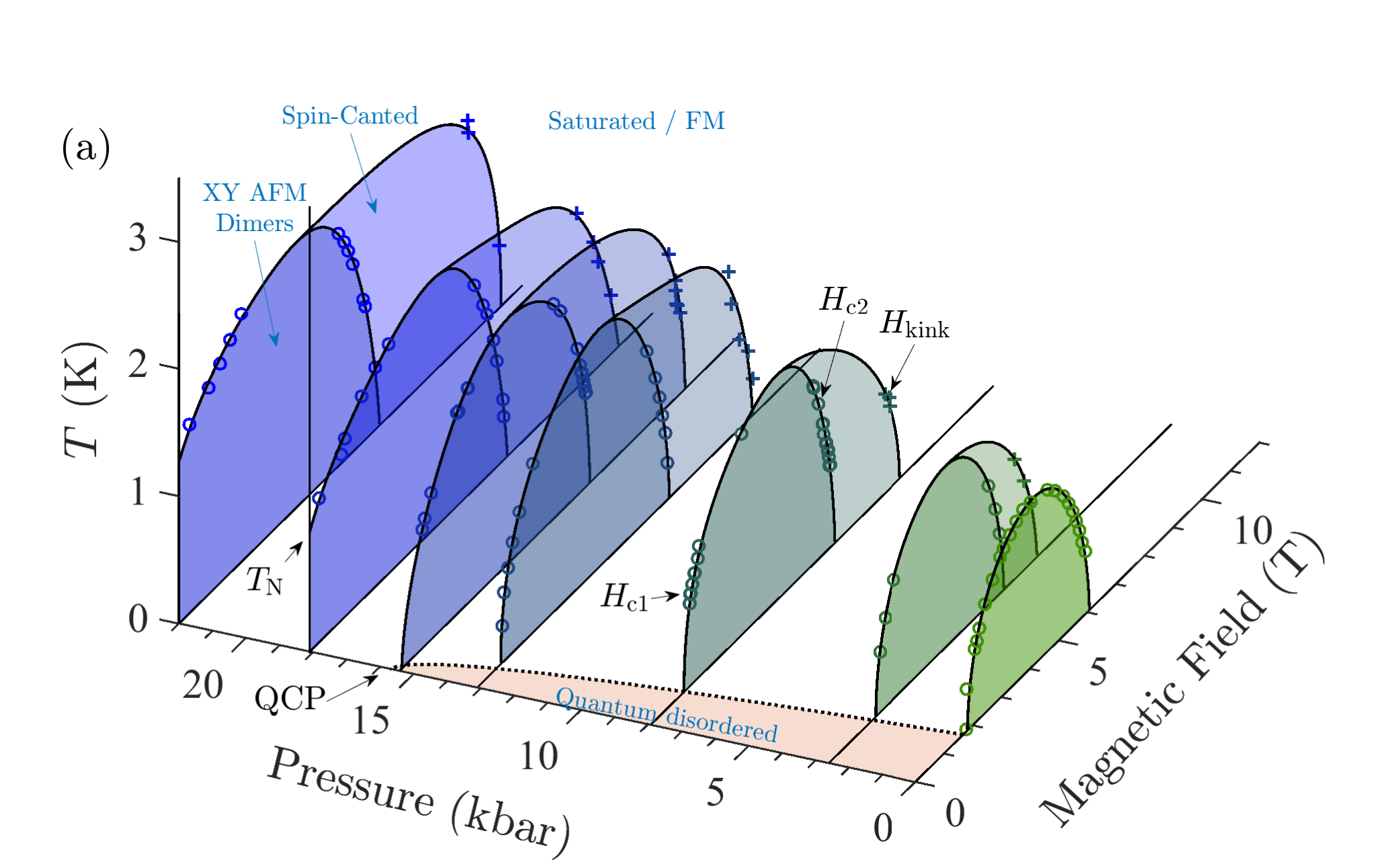}
\includegraphics[width= 0.32\linewidth]{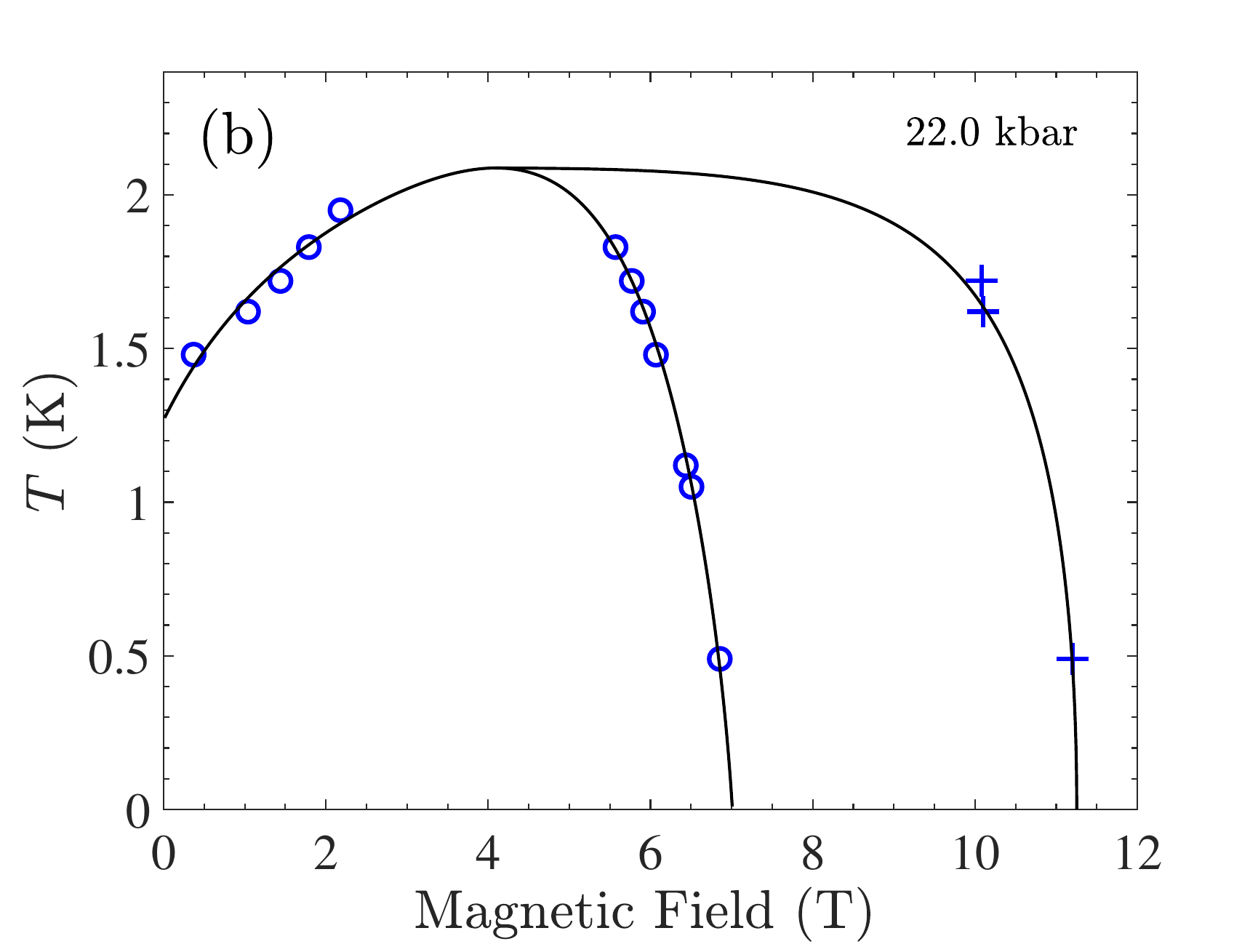}
\includegraphics[width= 0.32\linewidth]{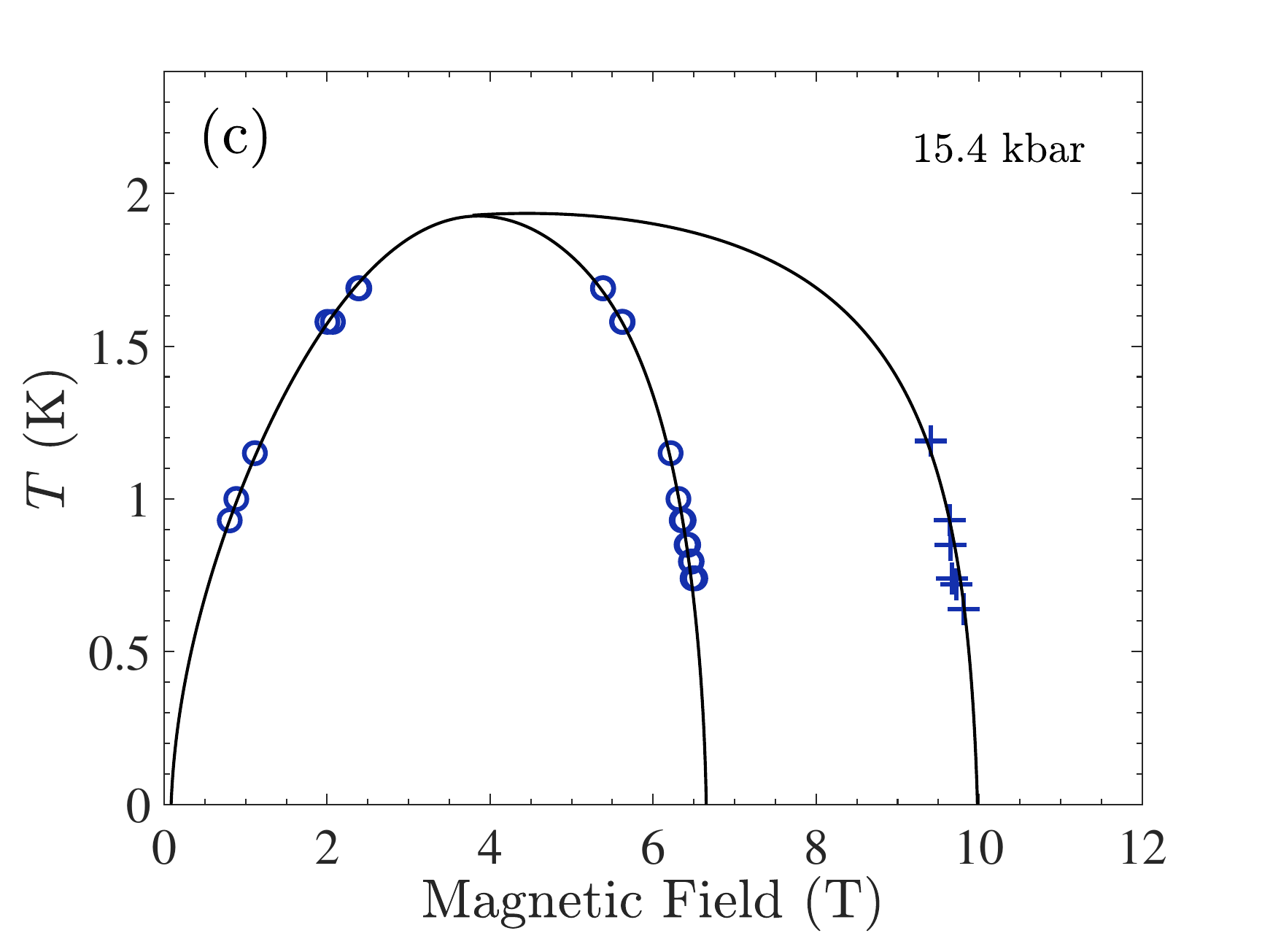}
\includegraphics[width= 0.32\linewidth]{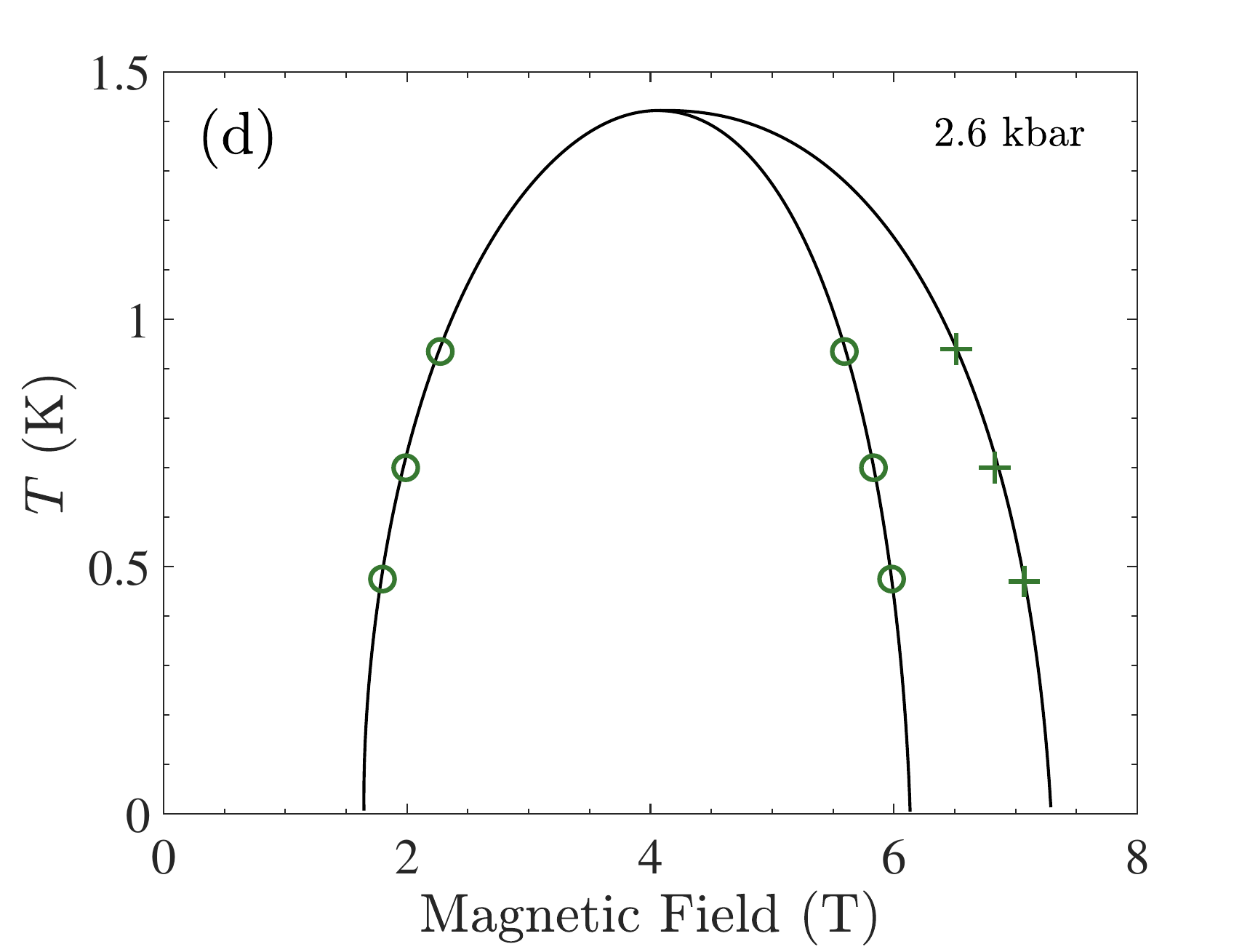}
\caption{ (a) Pressure evolution of the $H$-$T$ phase diagram for [Cu(pyz)$_{0.5}$(gly)]ClO$_4$ determined from the d$M$/d$H$ data, a subset of which is shown in Figure~\ref{Fig:CuCu_mag_pdep}. Darker shaded areas denote regions of AFM order enclosed by $H_{\rm c1}$ and $H_{\rm c2}$. Lighter shaded high-field regions enclosed by $H_{\rm c2}$ and $H_{\rm kink}$ correspond to the likely spin-canted regime discussed in the text. Dotted line and orange shaded region indicates the area of quantum disorder in the $p$-$H$ plane (shown again later in Figure~\ref{Fig:BPPhaseDia}). The square-root phase boundary intersects the pressure axis at a 15.7(5)\,kbar QCP. At higher pressures the system orders in the absence of field with a finite Néel temperature $T_{\rm{N}}$ as illustrated. (b,c,d) show the same 22.0, 15.4 and 2.6 kbar data plotted in part (a), but on 2D plots for more clarity.} \label{Fig:CuCu_BT_pdep}
\vspace{-0cm}
\end{figure*}

\subsection{Density functional theory}
Spin-polarized density functional theory (DFT) calculations were performed under the generalized gradient approximation (GGA)~\cite{Perdew1996} using the plane wave pseudopotential code, CASTEP~\cite{Clark2005}.
Calculations were converged to 1~meV per atom using a plane wave cutoff of 1500~eV and a 5$\times$5$\times$3 $k$-point grid~\cite{Monkhorst1976}.
The experimentally determined unit cell was allowed to relax, since we were unable to realise the spin configurations required to extract coupling constants from the experimental cell. On permitting the unit cell to relax, the cell volume increases by approximately 25\%. The increase is not uniform across all unit cell directions; we find increases of 3.1\%, 6.7\% and 16.9\% in the $a$, $b$ and $c$ directions respectively, which is expected due to the different bonding mechanisms present in these different directions. Specifically, the properties of the covalent bonds that form the corrugated sheets of dimers are more easily captured using GGA DFT than the weaker interdimer bonds and this has the effect of increasing the interdimer spacing. Using this DFT-relaxed cell we find all of the spin configurations required to extract $J_0$ and $J'_\mathrm{eff}$ are stable solutions, whereas considering the experimental cell we find that the magnetism is greatly suppressed. We therefore use the DFT-relaxed structure as the starting point for all subsequent calculations. From the DFT-relaxed unit cell, a series of calculations were performed by applying pressure isotropically over a range of 0--60~kbar. Further details of the DFT calculations are found in the appendix. 

\subsection{Quantum Monte Carlo calculations}
The Stochastic Series Expansion (SSE) quantum Monte Carlo (QMC) algorithm was used to simulate Hamiltonian (\ref{eq:Hamiltonian}) on finite size systems. The SSE is a finite temperature QMC based on  the stochastic sampling of the diagonal matrix elements of the density matrix expanded in a Taylor series using a suitable basis. An ``operator loop" cluster update, in conjunction with ``directed loop update" reduces the autocorrelation time for the system sizes studied here (up to $\sim 10^5$ spins) to a few Monte Carlo sweeps. It is known from earlier studies on weakly coupled Heisenberg chains that for spatially anisotropic systems, one needs to simulate rectangular lattices to obtain a monotonic system size dependence of calculated physical observables for extrapolating to the thermodynamic limit -- simulations of square lattices yield results that depend nonmonotonically on system size. Similar effects are expected for the present model and accordingly, lattices with dimensions $L_x\times L_x\times L_z$, were studied where $L_z$ is the dimension along the $c$-axis. An aspect ratio of $R=L_x/L_z=2$ was chosen for the finite size lattice studies with $8 \le L_x\le 32$. For each system size, an inverse temperature of $\beta J_0 = 8L_x$ was used to ensure that only the lowest energy state contributed to the calculated quantities. Estimates of the ground state properties were obtained from simultaneous finite-size and finite-temperature extrapolation of results from simulations on finite sized systems to the thermodynamic limit.

\section{Results}

\subsection{High-pressure magnetometry}

Fig.~\ref{Fig:CuCu_mag_pdep}(a) shows an example dataset, at a pressure of 13.1(5)\,kbar in this case, of the field dependence of the differential magnetization (d$M$/d$H$) for a single-crystal of [Cu(pyz)$_{0.5}$(gly)]ClO$_4$ measured at various temperatures using RF susceptometry. At the lowest measured temperature ($T = 0.3$\,K), two sharp cusps are seen in the data at $H_{\rm c1} = 0.8$\,T and $H_{\rm c2} = 6.7$\,T, that correspond to the first critical field at which the singlet-triplet energy gap is closed, and second critical field at which the spins are polarised along the field direction. Upon increasing temperature, $H_{\rm c1}$ and $H_{\rm c2}$ move in closer proximity to each other and cease to be resolvable as separate features for temperatures in the range $1.7 < T \leq 2.7$\,K. This behavior is qualitatively similar to that seen at ambient pressure~\cite{Brambleby2017}.

The pressure evolution of the d$M$/d$H$ data measured at fixed temperatures of 0.6(1)\,K is shown in Fig.~\ref{Fig:CuCu_mag_pdep}(b). $H_{\rm c1}$ and $H_{\rm c2}$ are seen to  shift to lower and higher fields, respectively, as pressure is applied. The cusp at $H_{\rm c1}$ also diminished in size as it is pushed to lower fields. By 18.1(5)\,kbar and above, $H_{\rm c1}$ has dropped to zero field, while the more dramatic cusp at $H_{\rm c2}$ rises to approximately 7\,T. At $p \geq 2.6(5)$\,kbar, an additional kink feature in d$M$/d$H$ emerges above $H_{\rm c2}$, marked in Fig.~\ref{Fig:CuCu_mag_pdep}(b) by $H_{\rm kink}$. This feature tracks to higher fields as pressure is increased up to the maximum value measured, 22.0(5)~kbar, separating further from $H_{\rm c2}$ at higher pressures.

\begin{figure*}[t!]
\centering
\includegraphics[width= 0.9\linewidth]{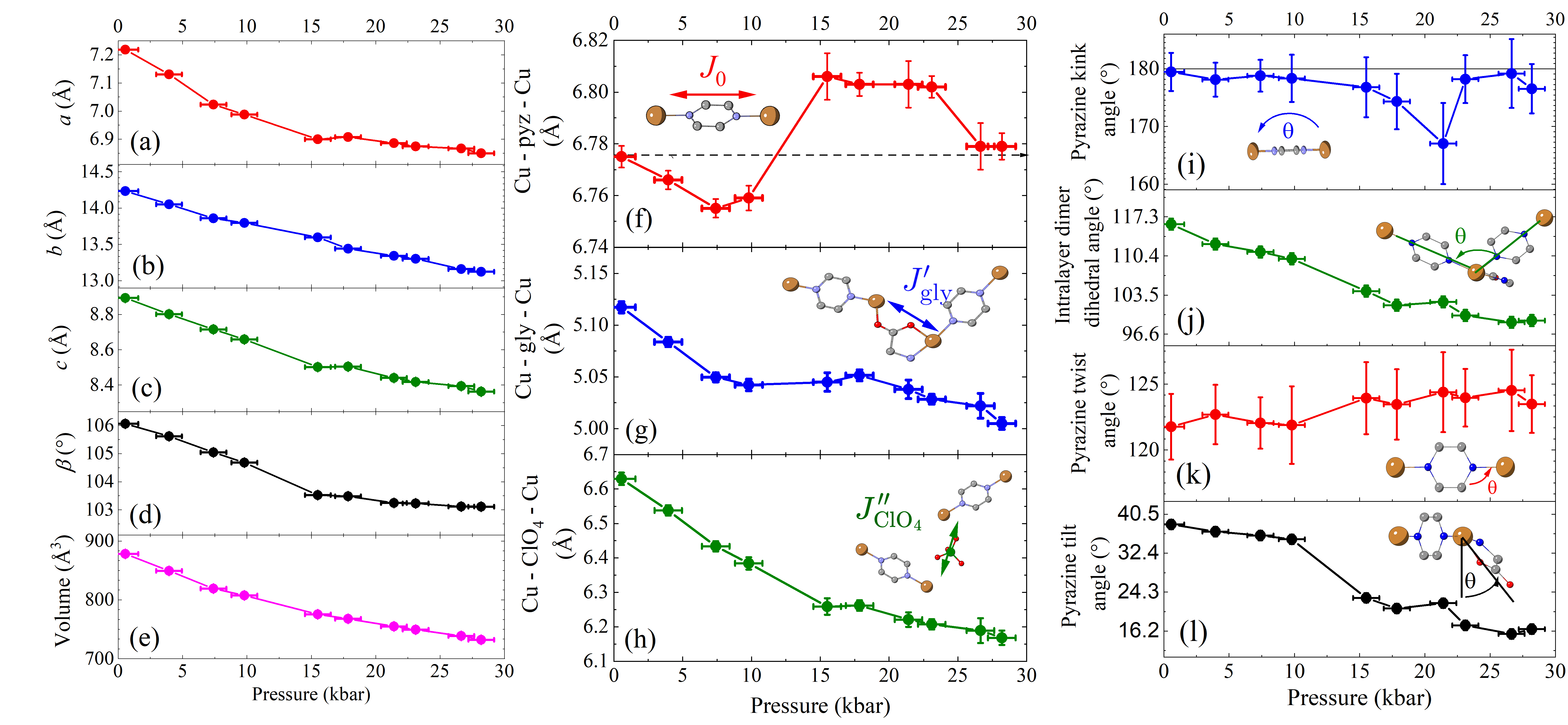}
\caption{Pressure dependences of various aspects of the crystal structure of [Cu(pyz)$_{0.5}$(gly)]ClO$_4$: (a-e) the unit-cell parameters; (f-h) the distances between adjacent Cu(II) ions; and (i-l) bond angles along the exchange pathways. The intralayer dimer dihedral angle (j) is measured between the N---Cu---Cu---N atoms, and the pyrazine tilt angle (l) is the dihedral angle between O---Cu---N---C atoms. Insets show cartoons of the exchange pathway or bond-angle being considered. Measurements were performed at 300~K and errors are plotted at 3$\sigma$.}  \label{Fig:CuCu_struc_pdep}
\vspace{-0cm}
\end{figure*}

By integrating the measured d$M$/d$H$ responses in Fig.~\ref{Fig:CuCu_mag_pdep}(b), one can extract the magnetisation $M$($H$) as shown in Fig.~\ref{Fig:CuCu_mag_pdep}(c). At $p \leq 13.1(5)$~kbar, the form of $M$($H$) remains typical for a dimer system, with a sharp upturn due to the closing of the singlet-triplet energy gap and a plateau at $H_{\rm c2}$ as the spins are nearly polarised along the field direction. For $p \geq 18.1(5)$~kbar, the abrupt low-field upturn in $M$($H$) is lost as $H_{\rm c1}$ is driven to zero field, and $M$($H$) rises with increasing gradient up to a sharp elbow at $H = H_{\rm c2}$. The data then rise more gradually before reaching saturation above $H = H_{\rm kink}$, best seen in the inset of Fig.~\ref{Fig:CuCu_mag_pdep}(c).

Extracting the critical fields $H_{\rm c1}(T)$, $H_{\rm c2}(T)$ and $H_{\rm kink}(T)$ at each pressure enables the temperature--field phase diagram to be constructed and shown in Fig.~\ref{Fig:CuCu_BT_pdep}. The ambient pressure dome of long-range $XY$ AFM order shown in green at ambient pressure changes in several ways as pressure is increased. \emph{First}, the dome grows, $H_{\rm c1}$ shifts to lower values of field and $H_{\rm c2}$ to higher, but the centre of the dome does not show any marked shift. Such behavior is indicative of an increase in the interdimer exchange interactions with increasing pressure, with the intradimer interaction remaining largely unchanged. This also explains the increase in the cut-off temperature at the top of the dome with applied pressure as an overall strengthening of the interdimer interactions leads to a stabilization of magnetic order to higher temperatures. \emph{Second}, an additional high-field phase between $H_{\rm c2}$ and $H_{\rm kink}$ emerges at finite pressures and grows in field range as pressure is increased.
\emph{Third}, $H_{\rm c1}$ is completely suppressed to zero field. The low-temperature values at each pressure are extrapolated to $T = 0$ and the resulting data are fitted to a square-root phase boundary~\cite{Rueegg2004}. The fit describes the data well, and is shown as the orange shaded region in the $p$-$H$ plane of Fig.~\ref{Fig:CuCu_BT_pdep}. In the absence of any signatures of first-order behavior or related changes in the phase diagram, we identify the presence of a QCP at a pressure of 15.7(5)\,kbar, which separates a high-pressure region of zero-field magnetic order from a quantum disordered region at lower fields. 
\emph{Finally}, the dome develops a clear asymmetry, the origin of which is considered in the Discussion section below.

\subsection{High-pressure crystallography}
\label{sec:CuCU_struc_Pdep}

To account for the changes in the magnetic phase diagram, it is extremely useful to study the pressure evolution of the crystal structure. To this end we carried out high-pressure single crystal X-ray diffraction at room temperature, the results of which are displayed in Fig.~\ref{Fig:CuCu_struc_pdep}. Panels (a-e) show how the unit-cell parameters change with pressure. The results show no symmetry-breaking phase transition or notable abrupt structural rearrangements. The cell volume decreases linearly across the full range, while the $\beta$ angle decreases monotonically up to 10(1)\,kbar. Beyond 10(1) to 15(1)\,kbar the rate of deformation decreases abruptly and the system becomes significantly more resistant to pressure.$\beta$ continues to decrease at a reduced rate up to the maximum pressure measured.

\begin{figure*}
  \includegraphics[width=1.6\columnwidth]{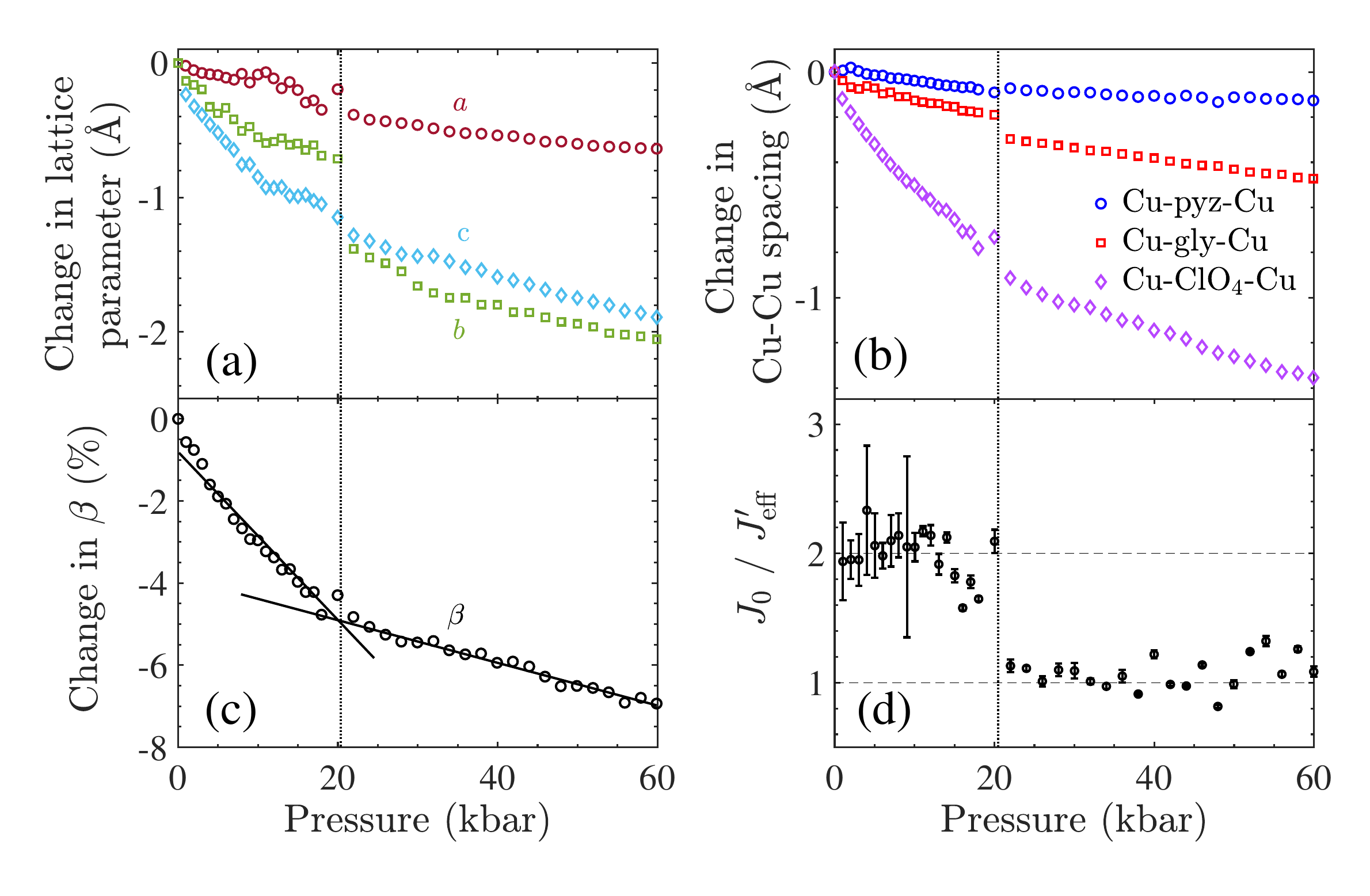}
  \caption{DFT results. (a) Theoretical evolution of lattice parameters of \CuPG\ with simulated external pressure. (b) Change in the distance between Cu ions along the available exchange pathways. (c) Pressure dependence of the unit cell angle $\beta$, lines have been added as a guide to the eye. (d) Ratio of coupling constants $J_0$ and $J'_{\mathrm{eff}}$ across the calculated pressure range, dotted lines indicate the two distinct regimes.}
  \label{Fig:DFT_panel}
\end{figure*}

Fig.~\ref{Fig:CuCu_struc_pdep}(f-l) highlight the effect of pressure upon various structural geometric parameters considered important to the magnetic exchange pathways. The data in Fig.~\ref{Fig:CuCu_struc_pdep}(f) show that the intradimer exchange pathway (which dictates the size of $J_0$) is relatively unchanged over the measured pressure range, with a total variation of about $\pm$0.03\,\AA{} across the entire pressure range. This can be compared to order-of-magnitudes larger variation seen across similar distances for the interdimer Cu-Cu distances shown in Fig.~\ref{Fig:CuCu_struc_pdep}(g) and ~\ref{Fig:CuCu_struc_pdep}(h) and discussed below. The change in kink-angle of the bridging pyrazine molecule is shown in Fig.~\ref{Fig:CuCu_struc_pdep}(i). A large kink can lead to an overlap of the $d_{xy}$ or $d_{yz}$ orbital of the Cu ion and the $\pi$-orbital on the pyrazine ring \cite{ButcherThesis}, resulting in a $\pi$ exchange mechanism occurring alongside the $\sigma$ exchange mechanism \cite{DosSantos2016} and increasing the overall exchange coupling in this direction. However, in our case the pyrazine molecule remains planar to a good degree of accuracy, with the exception of the point at 21(1)\,kbar. The twist angle of the pyrazine ring [Fig.~\ref{Fig:CuCu_struc_pdep}(k)] also shows no change within errors upon increasing pressure. While the pyrazine ring is seen to tilt significantly toward the glycine ligand on increasing pressure [Fig.~\ref{Fig:CuCu_struc_pdep}(l)], the pyrazine-tilt angle has been found to have little influence on the size of the magnetic exchange interactions in other Cu--pyz--Cu based low-dimensional magnets~\cite{Vela2013,Kenny2021}. Thus the pyrazine ring is relatively robust over the measured pressure range, and so we expect the primary intradimer exchange to remain roughly constant.

Fig.~\ref{Fig:CuCu_struc_pdep}(g) shows that on increasing pressure, the distance between adjacent Cu ions along the interdimer Cu--gly--Cu direction initially decreases, plateaus in the pressure range $10 \leq P \leq 17$\,kbar, then continues to decrease up to 28(1)~kbar; a total reduction in the intralayer distance between Cu ions on adjacent dimers of $\approx 0.12$\,\AA. This likely leads to an increase in the magnitude of the intralayer coupling ($J'_{\rm gly}$) between the dimers within the Q2D layers.

As shown in Fig.~\ref{Fig:CuCu_struc_pdep}(h), the most dramatic change is seen in the interlayer distance, where dimers are coupled via ClO$_4$ ions along the [10$\overline{1}$] direction. The interlayer Cu--Cu distance decreases linearly with increasing pressure up to 16~kbar, plateaus slightly up to 18~kbar, then continues to decrease up to 28~kbar. Over the measured pressure range, the interlayer distance between adjacent Cu ions is reduced by $\approx 0.5$\,\AA. 
For simplicity, the data shown are the through-space distance, however the atomic site separations along the expected interlayer exchange pathway show very similar behavior. This indicates the greatest effect pressure has on the system is to pack the dimer layers more tightly along the [10$\overline{1}$] direction, enhancing the interdimer exchange in this direction ($J''_{\rm ClO_4}$) and causing adjacent dimers within the layers to become more orthogonal to each other [see Fig.~\ref{Fig:CuCu_struc_pdep}(j)]. 

The observed stiffening of the $\beta$ angle and associated features in the room-temperature crystal structure is likely caused by the elimination of the most accessible void space at low pressures, followed by harder deformations resulting from continued compression. Although the stiffening occurs between 10 and 15\,kbar it is unlikely to be directly linked to the QCP seen at 15.7(5)\,kbar in the low-temperature magnetometry data. As discussed in more detail below and supported by quantum Monte-Carlo calculations, the QCP is instead driven by the gradual evolution of the exchange energies across the entire pressure range.

\subsection{Density Functional Theory}

To better understand the effects of pressure on the exchange couplings, we performed spin-polarized DFT calculations by applying pressure isotropically over a range of 0--60~kbar as described in the Methods section. We find that the spin configuration with the lowest total energy, and therefore the ground-state magnetic structure, is antiferromagnetic both within the Cu(II) dimers and between the dimers, this is consistent with both the intradimer $J_0$ and effective interdimer $J'_{\rm eff} = 4(J'_{\rm gly} +J''_{\rm ClO_4})$ coupling constants being antiferromagnetic. 

The calculated effects of externally applied pressure on \CuPG\ are shown in Fig.~\ref{Fig:DFT_panel}. All lattice parameters are predicted to decrease with applied pressure, with the decrease being continuous for $a$ and $c$. However there is a discontinuity in $b$ at approximately 20\,kbar [Fig.~\ref{Fig:DFT_panel}(a)]. It should be noted that the ambient-pressure unit cell calculated from DFT is approximately 25\% larger than is realised experimentally. The discontinuity in $b$ corresponds to a change in the unit cell angle $\beta$, and while $\beta$ always decreases with increasing pressure, the rate of decrease is predicted to slow at around 20~kbar, as seen in Fig.~\ref{Fig:DFT_panel}(c). A change in the orientation of ClO$_4$ molecules provides an explanation for the sudden changes in crystal structure seen by DFT at 20\,kbar with the ClO$_4$ molecule undergoing a rotation at 20\,kbar which allows for more efficient packing and the change in the $b$ axis shown in Fig.~\ref{Fig:DFT_panel}(c). Cu--Cu distances are shown in Fig.~\ref{Fig:DFT_panel}(b), where it can be seen that the most dramatic changes are predicted to occur between the Cu ions in the exchange pathway mediated by ClO$_4$ ($J_\mathrm{ClO_4}''$), where the Cu--Cu distance is expected to decrease by approximately 1~\AA~ with the application of 20\,kbar. We note that the abrupt changes in behavior predicted by DFT at 20\,kbar are not observed in the experimental data up to 28(1)\,kbar. On this point, it is worth noting that the experimental results show evidence for a degree of lability and dynamic disorder in the position of the ClO$_4$ anions that is not captured by DFT modelling and could explain any differences in the results between theory and experiment.

The effect of pressure on the exchange constants of \CuPG\ is also calculated and can be seen in Fig.~\ref{Fig:DFT_panel}(d).
At all pressures $J_0$ and $J'_{\mathrm{eff}}$ are found to be positive, indicating that the magnetic interactions are antiferromagnetic.
There are two distinct regimes for the coupling constants as with the structural parameters, with the coupling strengths changing from those found at ambient pressure above 20~kbar. Across the full range of pressures the coupling constant $J_0$ remains quite stable around $J_0= 8$~K, with a small increase at the highest pressures to $J_0\approx 9$~K. For the second coupling constant, $J'_{\mathrm{eff}}$, the behavior is more complex: in the region below 20~kbar the coupling is calculated to be roughly half the value of $J_0$, and  at higher pressures $J'_{\mathrm{eff}}$ increases, such that the ratio $J_0$/$J'_{\mathrm{eff}}~\approx 1$. In the region around the structural transition the effect on the coupling constants is less clear,  but the general trend shows that $J'_{\mathrm{eff}}$ decreases with increasing pressure over a range of $\approx 10$~kbar to its final value above 20~kbar.

\section{Discussion}

\subsection{The pressure-induced QCP}
The X-ray diffraction experiments under applied pressure (Fig.~\ref{Fig:CuCu_struc_pdep}) clearly indicate that the pyrazine molecules are robust to distortion, and the length of the Cu--pyz--Cu pathway, which mediates the primary intradimer exchange interaction ($J_{0}$), remains roughly constant across the whole pressure range measured. The Cu--glycine--Cu distance changes relatively smoothly by approximately 0.1\,\AA\ on increasing the pressure to 28(1)\,kbar. At ambient pressure, it is this pathway that mediates the main interdimer interaction ($J'_{\rm gly}$), and an increase in this exchange energy is therefore expected as pressure is applied. However, the largest change in the lattice occurs in the interlayer separation along $c$. This causes the Cu--ClO$_4$--Cu distance [Fig.~\ref{Fig:CuCu_struc_pdep}(h)] to reduce by close to 0.5\,\AA\ by the maximum applied pressure. Thus it is reasonable to assume that the interlayer exchange strength $J''_{\rm ClO_4}$, which was previously found to be negligible compared to $J_0$ and $J'_{\rm gly}$ at ambient pressure~\cite{Brambleby2017}, is the interaction that is most decisive in determining the behavior of the system under pressure. 

\begin{figure}[t!]
\centering
\includegraphics[width= \linewidth]{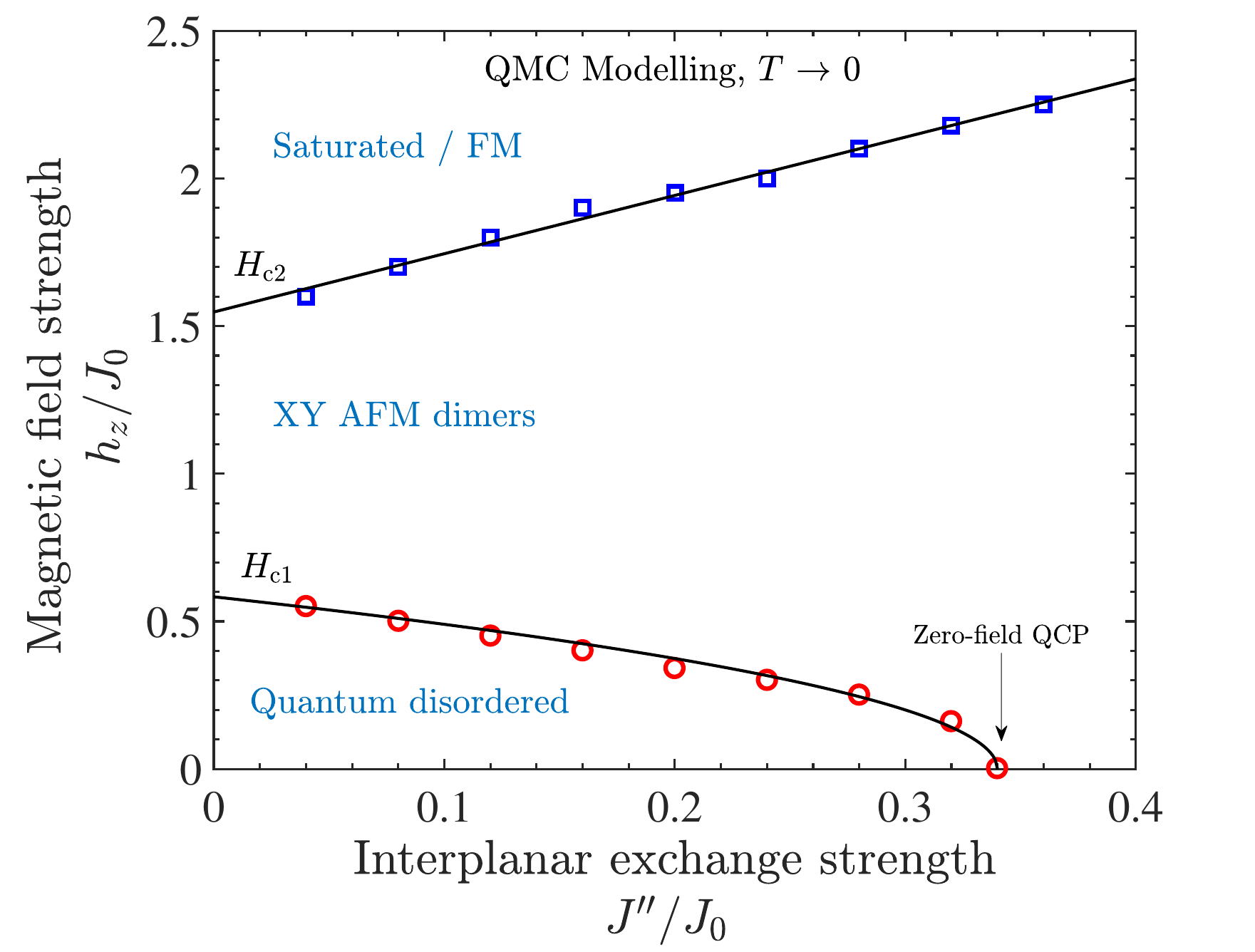}
\caption{Critical fields predicted from quantum Monte Carlo calculations as a function of changing the interplanar interdimer interaction. The ratio of inplane inter- and intradimer exchange $J'$/$J_0$ is set to 0.276, the value determined experimentally at ambient pressure~\cite{Brambleby2017}. The box size used was $N=16\times 16\times 8$. As argued in the text, the effect of pressure can be primarily linked to an increase in $J''$, and so these calculations can be compared to the experimental field-pressure phase diagram of Figure~\ref{Fig:BPPhaseDia}.}  
\label{Fig:QMC_results}
\vspace{-0cm}
\end{figure}

This conclusion is supported by our DFT calculations (Fig.~\ref{Fig:DFT_panel}), which predict that the Cu--ClO$_4$--Cu distance should decrease significantly under pressure, while the Cu--glycine--Cu pathway is much less compressed, and the Cu--pyz--Cu is the most robust to change. The calculations show further that the effective interdimer coupling, the sum of all interdimer exchange strengths, doubles relative to the intradimer coupling across 30\,kbar of applied pressure. Although this result is influenced by the quite abrupt change predicted to occur in the $b$-axis lattice parameter at 20\,kbar (which is not seen in the experimental data), the reduction in interlayer distance clearly  plays a key role. 

\begin{figure}[t!]
\centering
\includegraphics[width= \linewidth]{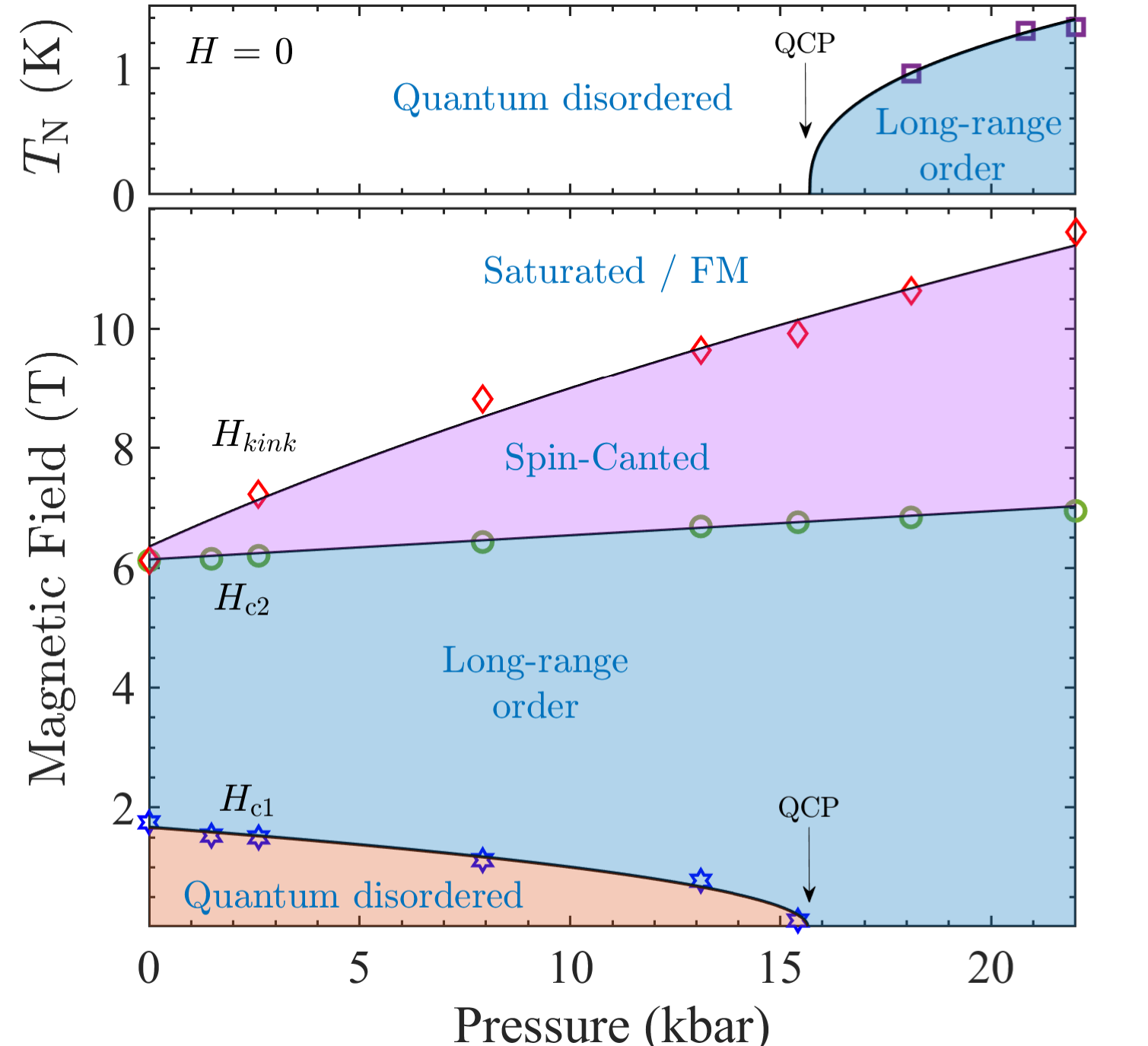}
\caption{Lower panel: $p$-$H$ phase diagram at low temperature, extrapolated from the data shown in Fig.~\ref{Fig:CuCu_BT_pdep}. $H_{\mathrm{c}1}$is suppressed to zero-field and zero-temperature at $p_{\mathrm{c}}$= 15.7(5)\,kbar. Upper panel: pressure evolution of the zero-field dimer ordering temperature $T_{\mathrm{N}}$ at pressures above $p_{\mathrm{c}}$. Solid lines are guides to the eye.}  \label{Fig:BPPhaseDia}
\vspace{-0cm}
\end{figure}

To confirm the origin of the zero-field ordered state, QMC calculations were also performed as outlined in the Methods section.  The results shown in Fig.~\ref{Fig:QMC_results} indicate that $H_{\rm c1}$, the critical field that separates $XY$ order from quantum disordered phase can be pushed to zero by an increase from 0 to 0.34 in the $J''_{\rm ClO_4}$/$J_0$ ratio when $J'_{\rm gly}$ is fixed at its experimentally determined ambient-pressure value. This would suggest that the simplest explanation for the observed QCP at 15.7(5)\,kbar and the associated onset of zero-field magnetic order, is the increase under pressure of the size of the interlayer interdimer coupling mediated through the perchlorate molecules, which in turn is driven by the anisotropic compression of the crystal lattice. 
Previous heat capacity measurements have shown that the magnetic properties of \CuPG\ are highly two-dimensional at ambient pressure~\cite{Brambleby2017}. Our present results therefore imply that the primary driver of the zero-field magnetic order is a dimensional crossover from Q2D to three-dimensional magnetism. The Q2D spin-1/2 dimer material PHCC also shows a pressure-induced QCP. There the effect of pressure is to weaken the intradimer coupling and push the dome of magnetic order to lower fields~\cite{Perren2015}, which is different from the broadening of the dome we observe in our material.

\subsection{Quantum fluctuations and asymmetry in the phase diagram}

The magnetometry measurements point to a somewhat more complicated situation than the simple picture above. Fig.~\ref{Fig:BPPhaseDia} shows how the three critical fields observed in the dynamic susceptibility measurements evolve as a function of pressure. $H_{\rm c1}$ is the position of the lower-field cusp in the susceptibility, or the point at which the magnetization begins to rise quickly with field (see Fig.~\ref{Fig:CuCu_mag_pdep}). It marks the transition between the quantum-disordered dimer phase and the $XY$-ordered state and is suppressed by initial pressurization at a constant rate of -71(4)\,mT/kbar up to 13\,kbar, after which it drops abruptly to the QCP at 15.7(5)\,kbar.  

The field $H_{\rm c2}$ is the location of the large cusp in susceptibility, which marks the point at which the magnetization begins to plateau. At ambient pressure, this corresponds to the transition between the $XY$ ordered state and the field-saturated phase, where the Zeeman energy surpasses the sum of all the AFM interactions between the spins. As pressure is applied, $H_{\rm c2}$ rises linearly at 40(1)\,mT/kbar, which is notably slower than the pressure-induced drop in $H_{\rm c1}$.  

That $H_{\rm c1}$ and $H_{\rm c2}$ vary with pressure at different rates is not straightforward to understand. It has previously been noted that the effect of quantum fluctuations is greater on the lower-critical-field transition, where the bosonic effective mass is strongly renormalized, than for the transition at $H_{\rm c2}$~\citep{Brambleby2017}. This is manifested by a strong asymmetry in the entropy change seen at the two transitions as measured using heat capacity \citep{Brambleby2017}. It has been predicted that the disparity in the sizes of the peaks in heat capacity at the two transitions should increase strongly as $H_{\rm c1}/H_{\rm c2}$ decreases, diverging as this ratio tends to zero~\cite{Kohama2011}. This effect is clearly reflected in the sizes of the features seen at the two field-induced transitions in our susceptibility data, as shown in Fig.~\ref{Fig:CuCu_mag_pdep}(b). As the pressure increases, the size of the feature at $H_{\rm c1}$ diminishes quickly, indicating that the renormalization due to quantum fluctuations gets significantly stronger as this transition is pushed to lower fields. Finally, as the pressure-induced QCP is approached the feature at $H_{\rm c1}$ disappears altogether, in keeping with the predictions of Reference~\cite{Kohama2011}.

In principle, the variation in the relative importance of quantum fluctuations at the two transitions could also explain their different shifts as a function of pressure. However, the QMC calculations are also sensitive to effects of quantum fluctuations and the results show the two transitions evolving at very similar rates. 

This points to an unconventional asymmetry in the phase diagram beyond that which might be expected from the effect of quantum fluctuations alone. A possible explanation could arise if the size of the effective interlayer coupling $J''_{\rm ClO_4}$ depends on the component of the spins perpendicular to the $c$-axis. This would lend a field dependence to the interdimer exchange strength, which would diminish as the spins become more aligned with $c$. Our magnetometry measurements were performed with $H\parallel [011]$ and so, if this proposition has merit, then the interlayer coupling would become less effective with increasing field and hence the transition at $H_{\rm c1}$ would feel the pressure-induced enhancement of $J''_{\rm ClO_4}$ more strongly than the $H_{\rm c2}$ transition, giving rise to the observed difference in pressure evolution of the two transitions.

\subsection{High-field phase and the DM interaction}
In addition to $H_{\rm c1}$ and $H_{\rm c2}$, for pressures above ambient another higher-field feature is seen at $H_{\rm kink}$, marking a very small step in magnetisation (as shown in the inset to Fig.~\ref{Fig:CuCu_mag_pdep}(c)). Assuming that it arises from the bulk behavior of the system, this feature suggests that, for elevated pressures, the spins in the state above $H_{\rm c2}$ are canted slightly from the field direction, and fully align parallel to $H$ only for $H > H_{\rm kink}$. Such a canting is typically caused by energy scales in the spin Hamiltonian, such as anisotropy, that compete with the Zeeman term. The size of the step in magnetisation that occurs at $H_{\rm kink}$ increases with pressure, and at 23~kbar is approximately 2\% of the saturated moment. Assuming all spins are canted at the same angle from the field, this would correspond to a canting angle of about $11.5^\circ$. The crossover field $H_{\rm kink}$ itself increases rapidly with pressure at a rate of 230(10)\,mT/kbar, which is significantly faster than the changes in either $H_{\rm c1}$ or $H_{\rm c2}$. 

The non-centrosymmetric structure of \CuPG, with its alternating orientation of dimers within the $ab$-plane, allows for the existence of an antisymmetric DM interaction between spins, which is known to give rise to canted spin structures. No evidence for the effect of such an interaction was forthcoming in the previous ambient pressure studies of this system. (DM interactions result from a second-order correction to the energy arising from the combination of spin-orbit interaction and exchange, so are expected to be smaller than Zeeman, single-ion anisotropy or exchange effects). However, as can be seen from Fig.~\ref{Fig:CuCu_struc_pdep}(j), our X-ray data suggest that the dihedral angle between adjacent dimers in the corrugated planes sharpens from $116.0(8)^\circ$ at ambient pressure to $99.0(9)^\circ$ by 28(1)\,kbar, potentially increasing the strength of the DM interaction, and thus providing an explanation for the enhancement of the canting angle and the field $H_{\rm kink}$ as pressure grows. 

We might expect any DM interaction present to influence the transition at $H_{\rm c1}$ as well as that at $H_{\rm c2}$. Indeed, the energetic contribution from the DM interaction should diminish quickly with increasing applied field as the relative angle between neighboring spins is reduced, and so the perturbation to the Hamiltonian from DM is likely larger in the low-field limit. Exactly how this $H_{\rm c1}$ transition between quantum-disordered and $XY$-ordered states would be affected by a DM term is hard to anticipate without detailed calculations and a better knowledge of the direction and size of the dominant DM vector. However, the combination of DM and the field-dependent interlayer coupling proposed above would seem to provide the most plausible explanation for the small realignment of the spins at $H_{\rm kink}$, as well as the asymmetry between low-field and high-field transitions via an intrinsic, bulk mechanism, without invoking some local effect.

\section{Summary}
Given the discussion above, it seems clear that the interlayer exchange term increases strongly with pressure and probably drives the quantum phase transition observed in the magnetometry data at 15.7(5)\,kbar. Nevertheless, accounting for the pressure dependence of all of the features seen in the susceptibility of \CuPG\ is not trivial. The observed results likely arise from both symmetric and antisymmetric interaction terms in the Hamiltonian, and a field dependence in one or more of these interactions is possible. Inelastic neutron scattering would help to shed further light on exactly what drives the phase transitions and efforts are underway to grow sufficiently sized deuterated crystals to enable these measurements. With its experimentally accessible exchange energies, this is one of very few $S = 1/2$ dimer materials on which neutron scattering studies could be performed across both field-induced QCPs. In the meantime, it is apparent that the molecular bridges that connect the spins in this class of material support an interesting hierarchy of competing energy scales. Furthermore, the resulting crystal structure undergoes anisotropic compression on application of pressure and the balance of these interactions seems to evolve in a non-trivial manner both in pressure and magnetic field.

\appendix*

\section{Details of density functional theory calculations}

\begin{figure}[t!]
  \includegraphics[width=\columnwidth]{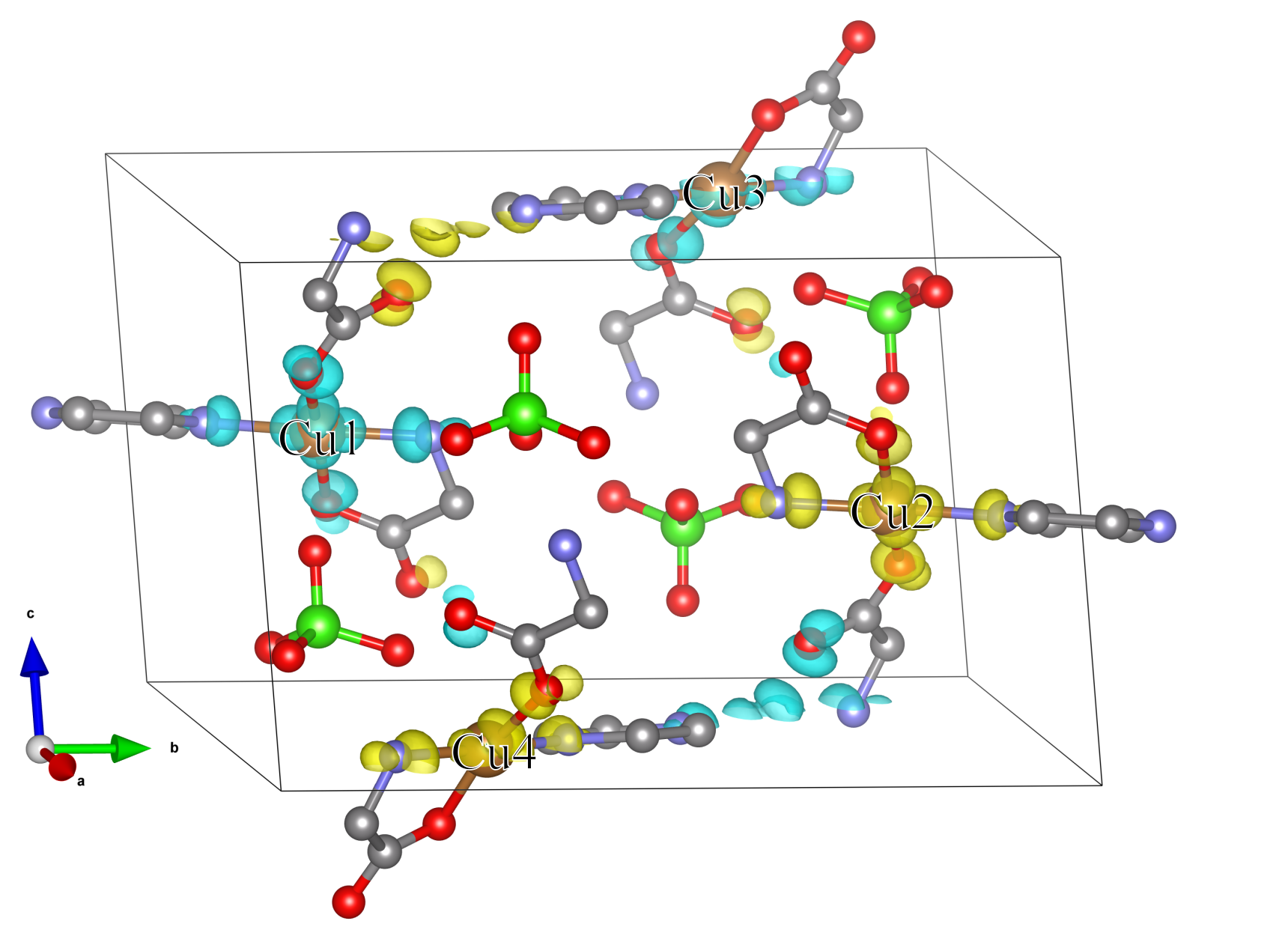}
  \caption{Spin density of the antiferromagnetic ground state of \CuPG. Labels on the Cu ions refer to the order of ions considered in each different spin configuration. H atoms have been removed for clarity}
  \label{Fig:spin_label}
\end{figure}

\begin{table}[t!]
  \begin{center}
    \begin{tabular}{ l r }
      \hline
      Spin configuration  & Energy \\      
      \hline
      $\uparrow$$\downarrow$$\uparrow$$\uparrow$ &  $E_0$  \\
      $\uparrow$$\downarrow$$\downarrow$$\uparrow$ &  $2J_0 + J'_{\mathrm{eff}} + E_0$ \\
      $\uparrow$$\downarrow$$\downarrow$$\downarrow$ &  $E_0$ \\
      $\uparrow$$\uparrow$$\downarrow$$\uparrow$ &  $E_0$ \\
      $\uparrow$$\uparrow$$\uparrow$$\downarrow$ &  $E_0$ \\
      $\uparrow$$\downarrow$$\uparrow$$\downarrow$ &  $ 2J_0 - J'_{\mathrm{eff}} + E_0$  \\
      $\uparrow$$\uparrow$$\downarrow$$\downarrow$ &  $-2J_0 + J'_{\mathrm{eff}} + E_0$\\
      \hline
    \end{tabular}
  \end{center}
  \caption{Energies of configurations of spins in \CuPG.}
  \label{Tab:energies}
\end{table}

To calculate the coupling constants, a single unit cell containing four
Cu(II) ions was used, which allows for a maximum of eight unique magnetic configurations.
Using a single unit cell then allows us to extract $J_0$ and
$J'_{\mathrm{eff}}$, where
$J'_{\mathrm{eff}}=4(J_\mathrm{pyz}'+J_\mathrm{ClO_4}'')$.

A supercell would be required to extract all three coupling constants, however this would be prohibitively large computationally.
For each pressure, calculations were performed by initialising 7 different spin configurations and calculating the total energy of each.
These total energies are then used to solve a set of simultaneous
equations~\cite{Thomas2017} in terms of $J_0$ and $J'_{\mathrm{eff}}$. 

As the calculations provide an over-complete set, spin configurations that are equivalent in our model can be used to estimate the uncertainty in the couplings.

Fig.~\ref{Fig:spin_label} shows isosurfaces of the ground-state spin density in \CuPG\, there is significant spin density on each of the Cu(II) ions with neighbouring ions having opposite signs. In each case the Cu(II) ions induce spin density on the O atoms in the glycine ligands and the N atoms in pyrazine. Table.~\ref{Tab:energies} shows the system of equations used to
evaluate the exchange couplings in \CuPG. The states are represented by the direction of the magnetic moment on each Cu ion in the order Cu1 Cu2 Cu3 Cu4 as given in Fig.~\ref{Fig:spin_label}. Each equation in the system contains the total energy of system excluding magnetic exchange effects, $E_0$, in many of the spin configurations the contributions to $J_0$ and $J'_{\mathrm{eff}}$ effective cancel leaving only the contribution from $E_0$.

\begin{acknowledgments}
We are indebted to the late Jamie Manson for instigating this work, for his role in designing and growing the samples and for many other invaluable contributions. We thank T. Orton and P. Ruddy for technical assistance.
This project has received funding from the European Research Council (ERC) under the European Union\textquoteright s Horizon 2020 research and innovation programme (grant agreement No. 681260). A portion of this work was performed at the National High Magnetic Field Laboratory, which is supported by National Science Foundation Cooperative Agreement No. DMR-1644779 and the State of Florida. We acknowledge the support of EPSRC (UK) [grant no. EP/N032128/1] and Durham Hamilton HPC. This work used the ARCHER2 UK National Supercomputing Service (https://www.archer2.ac.uk) [grant no. EP/X035891/1].
For the purpose of open access, the author has applied a Creative Commons Attribution (CC-BY) licence to any Author Accepted Manuscript version arising from this submission. Data presented in this paper will be made available at wrap.warwick.ac.uk/180953. Complete crystal information .CIF files, including embedded structure factors and SHELX refinement instructions (.res), are made available via. the CCDC, Deposition Numbers CCD 2302931-2302941.
\end{acknowledgments}

\end{document}